\documentclass[manuscript]{aastex}
\usepackage{ulem}

\shorttitle{} 
\shortauthors{Ishikawa \& Tsuneta}

\begin{document}
\title{Relationship between vertical and horizontal magnetic fields in the quiet Sun}
\author{Ryohko Ishikawa \altaffilmark{1,2} \and Saku Tsuneta\altaffilmark{2}}
\email{ryoko.ishikawa@nao.ac.jp}

\altaffiltext{1}{Department of Astronomy, University of Tokyo, Hongo,
Bunkyo-ku, Tokyo 113-0033, Japan} \altaffiltext{2}{National Astronomical
Observatory of Japan, 2-21-1 Osawa, Mitaka, Tokyo 181-8588, Japan}

\begin{abstract}
Vertical magnetic fields have been known to exist in the internetwork region for decades,
while the properties of horizontal magnetic fields have recently been 
extensively investigated with \textit{Hinode}.
Vertical and horizontal magnetic fields in the internetwork region
are considered to be separate entities and
have thus far not been investigated in a unified way.
We discover clear positional association between the vertical and horizontal magnetic fields in the internetwork region with \textit{Hinode}.
Essentially all of the horizontal magnetic patches
are associated with the vertical magnetic patches.
Alternatively, half of the vertical magnetic patches accommodate the horizontal magnetic patches.  
These horizontal patches are located around the borders of the vertical patches. 
The intrinsic magnetic field strength as obtained with the Stokes $V$ line ratio inside the horizontal
patches is weak, and is in sub-equipartition field regime ($B<700$~G), while
the field strength outside the horizontal patches ranges from weak to strong (kG) fields. 
Vertical magnetic patches are known to be concentrated on mesogranular and  supergranular boundaries,
while the horizontal magnetic patches are found only on the mesogranular boundaries.
These observations provide us with new information on 
the origin of the vertical and horizontal internetwork magnetic fields, in a unified way.
We conjecture that internetwork magnetic fields are 
provided by emergence of small-scale flux tubes with bipolar footpoints,
and the vertical magnetic fields of the footpoints are intensified to kG fields due to convective collapse.
Resultant strong vertical fields are advected by the supergranular flow, and eventually form the network fields. 
\end{abstract}

\section{Introduction}
\label{intro}
In the quiet regions of the Sun, strong vertical fields are concentrated around the boundaries of
the supergranulation,
and are referred to as network fields.
The intrinsic magnetic field strength is determined to be 1--2~kG \citep{Stenflo1973SoPh},
and is comparable to the field strength in the plage regions.
The region inside the network is referred to as the internetwork region.
\citet{Livingston1971} found that  
the internetwork region has weak magnetic fields.
Since then, spectropolarimetric observations have been carried out 
to explore the internetwork magnetic fields 
\citep[e.g.,][]{Lin1995ApJ,Lites1996ApJ,Lin1999ApJ,Socas2004ApJ}.
Most of the observations have used only a circular polarization signal, 
and thus, the magnetic field inclination could not be deduced.
\citet{Khomenko2003AA} carried out the full spectropolarimetric observation 
using \ion{Fe}{1} 1565~nm lines, which have higher Zeeman sensitivity \citep{Lin1995ApJ}
than the often used \ion{Fe}{1} 630.2~nm lines,
and found that the magnetic fields in the internetwork region have a broad range of inclinations
with peak essentially at vertical.

\citet{Schrijver1997ApJ} concluded that all of the flux in the network is 
generated locally \citep[e.g,][]{Martin1990},
and only the small net flux density is the result of dispersal from decaying active regions.
They also point out that the flux supplied by the emergence of the ephemeral regions 
is the most important source for the quiet-Sun network and that they cannot put constraints
on the amount of flux that is injected on scales smaller than that of the ephemeral regions.
Even with these pioneering works, the properties and the origin of the internetwork fields have been far from understood,
and high-resolution observations of the internetwork region is critically
important to clarify the generation mechanism of the quiet Sun magnetic fields.

The Solar Optical Telescope spectropolarimeter \citep[SOT/SP;][]
{Tsuneta2008Soph,Suematsu2008Soph,Ichimoto2008Soph,Shimizu2008Soph} 
on board  \textit{Hinode} \citep{Kosugi2007Soph}  
has allowed us to measure the full Stokes spectra with high polarization sensitivity
and with high spatial resolution.
The SOT/SP revealed the ubiquitous horizontally inclined magnetic fields 
with the intrinsic field strength of a few hundred G in the internetwork regions
\citep{Orozco2007PASJ,Orozco2007ApJL,Lites2008ApJ}. 
Such internetwork horizontal magnetic fields 
were originally reported by \citet{Lites1996ApJ} and \citet{DePontieu2002ApJ}.
Owing to the high spatial resolution and
the high polarimetric sensitivity of SOT/SP,
the properties of horizontal fields have been studied in greater detail
\citep[e.g.,][]{Centeno2007ApJL,Ishikawa2008AA,Jin2009ApJ,Martinez2009ApJ,Ishikawa2010ApJ}.
The horizontal magnetic fields in the internetwork region were also confirmed and investigated
 \citep{Danilovic2010ApJL} with the recent Sunrise/IMaX mission 
 \citep{Barthol2010,Solanki2010ApJ,Martinez2010}.
The horizontal magnetic fields are highly transient with the lifetime raging from 1 to 10 minutes
\citep{Ishikawa2009AA}.
These horizontal magnetic fields are observed in the quiet Sun as well as 
in a weak plage region \citep{Ishikawa2008AA} 
and a polar region \citep{Tsuneta2008ApJ}.
This ubiquity is consistent with that of seething horizontal magnetic fields 
reported earlier by \citet{Harvey2007ApJL}
using the SOLIS and GONG instruments.
The discovery of the ubiquitous horizontal fields 
provides us with new insight to better understand 
the quiet Sun magnetism in more general perspective. 

An important question on the properties of the internetwork magnetic fields is
concerned with their inclination distributions.
\citet{Orozco2007ApJL} and \citet{Lites2008ApJ}
reported that the horizontal magnetic fields are dominant in the internetwork region,
while \citet{Beck2009AA} and  \citet{Khomenko2003AA} found the dominance of 
the vertical magnetic fields.
Some authors pointed out that magnetic field vectors  are randomly oriented 
and there is no preference of a specific inclination in the internetwork region 
\citep{Martinez2008AA,Asensio2009ApJ,Bommier2009AA}.
Most of these authors exploit 
the probability density function (PDF) of the inclination angles and the field strengths.
Such PDFs represent one aspect of the internetwork magnetic fields.
We, however, stress that the original spectropolarimetric images 
have rich information which cannot be represented by simple PDF.
For instance, one can address, based on these images, 
whether the horizontal field patches are somehow associated with the vertical fields
or are a completely different entity from the vertical ones.
In this paper, we study the spatial distributions of the vertical and horizontal magnetic fields 
in an internetwork region, and
aim to address a question on how these horizontal and vertical fields are 
spatially distributed and possibly related.

We emphasize the importance of the spatial distribution of
the vertical and the horizontal magnetic fields in terms of the convective patterns with different scale sizes 
to understand the relationship of the magnetic fields with the convective motion.
\citet{Ishikawa2010ApJL} found that the horizontal fields appear 
with clear cellular pattern of a typical
scale of 5-$10\arcsec$ and that the cellular structure coincides in position
with the negative divergence of the horizontal flow field, i.e.,
mesogranular boundaries with downflows. 
As to the internetwork vertical fields, similar concentrations have been observed 
\citep{Dominguez2003AA}.
Mesogranular flows would play an important role on the
generation and distributions of the internetwork magnetic fields.
A natural question is whether there are similar concentrations of
the internetwork fields on the supergranular scale.
In this paper, we present the spatial distributions of the vertical and the horizontal
magnetic fields in terms of mesogranular and supergranular convection.

One of the major issues about the internetwork fields is their origin.
One possibility is the ``recycling" mechanism pointed out by \citet{Ploner2001}.
The remnants of active region flux are continually recycled by granular and supergranular convection
to form the network as well as internetwork fields.
Another possibility is that local dynamo process as suggested by \citet{Cattaneo1999ApJL}
works for the production and maintenance of the internetwork fields.
An ``explosion" of a rising flux tube from deeper convection zones \citep{Moreno1995ApJ}
may generate weak magnetic fields in the upper convection zone 
that would be the source of the quiet-Sun magnetic fields.

The field strength of the horizontal fields in the internetwork region is a few hundred G, 
which is below the equipartition field strength corresponding to the granular convective motion. 
The histograms (PDF) of the intrinsic magnetic field strengths of the horizontal fields
are exactly the same in the polar region, the quiet sun, and a weak
plage region in spite of the significant difference in the amount of the vertical magnetic flux  
\citep{Ishikawa2009AA,Ito2010ApJ}.
This ubiquity combined with the unique PDF and their sub-equipartition field strength
implies that the horizontal magnetic fields are generated by the local dynamo process,
in which the granular convective motion amplifies the magnetic fields by folding and stretching. 
By simultaneously focusing on both horizontal and vertical fields, we try to reveal the origin of the internetwork 
magnetic fields in a unified way in this paper.

\section{Observations and Data Reduction}
\subsection{\textit{Hinode} Observations}
\label{obs}
\begin{figure*}
%\epsscale{1.0}
\epsscale{0.4}
\plotone{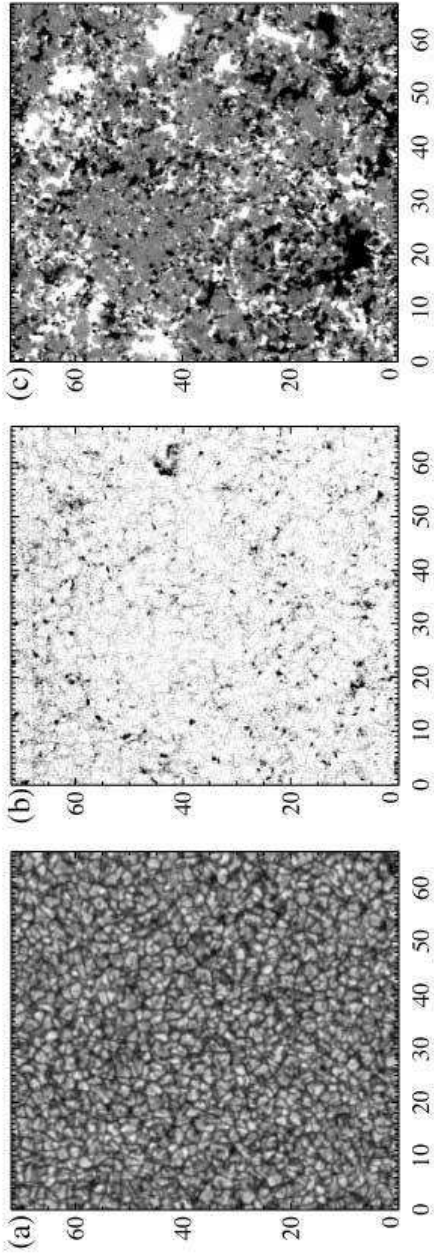}
\caption{A $67\farcs5\times72\arcsec$ sub-field selected for analysis.
(a) continuum intensity map (b) map of net linear polarization $L_{\rm tot}$ as defined by equation (\ref{eq:Ltot}).
(c) map of circular polarization $V_{\rm tot}$ as defined by equation (\ref{eq:Vtot}).}
\label{fig_region}
\end{figure*}

A quiet region with size of 302\arcsec $\times$ $162\arcsec$  at the disk center was 
observed using the Solar Optical Telescope Spectropolarimeter 
(SOT/SP) on March 10, 2007. The SOT/SP provides us 
with the full Stokes $I$, $Q$, $U$ and $V$ spectra covering the \ion{Fe}{1} 630.15~nm 
(Land\'e factor $g_{6302}=1.667$) and \ion{Fe}{1} 630.25~nm ($g_{6301}=2.50$) lines
with the wavelength sampling of 2.14~pm \citep{Ishikawa2010ApJ}.
Each slit position has an exposure time of 4.8~s.
The noise level $\sigma_{0}$ estimated from the continuum 
in the Stokes $Q$ and $U$ spectra is $1.2\times10^{-3}I_{c}$, and 
that in the Stokes $V$ spectrum is $1.1\times10^{-3}I_{c}$ for a single wavelength pixel, 
where $I_{c}$ is the continuum intensity.
The scanning step is 0\farcs15 and the spatial  sampling along the slit is 0\farcs16.
The dark current, the flat field corrections, and the polarimetric calibration were 
performed using ``sp\_prep" software available in SolarSoft.
The identical data set was extensively used by many authors \citep[e.g.,][]{Orozco2007PASJ,Orozco2007ApJL,Lites2008ApJ,Asensio2009ApJ,Viticchie2010AA} to explore the quiet Sun magnetism.
In this paper we analyze a $67\farcs5\times72\arcsec$ subfield, which covers  one supergranular cell,
and contains both network and internetwork regions (Figure~\ref{fig_region}) .    
Approximate scan time of the subfield is 36~min, which is much shorter than
the lifetime of the supergranular cell.

\subsection{Selection of valid pixels}
\label{pix}
\begin{figure*}
%\epsscale{2}
\epsscale{1.0}
\plotone{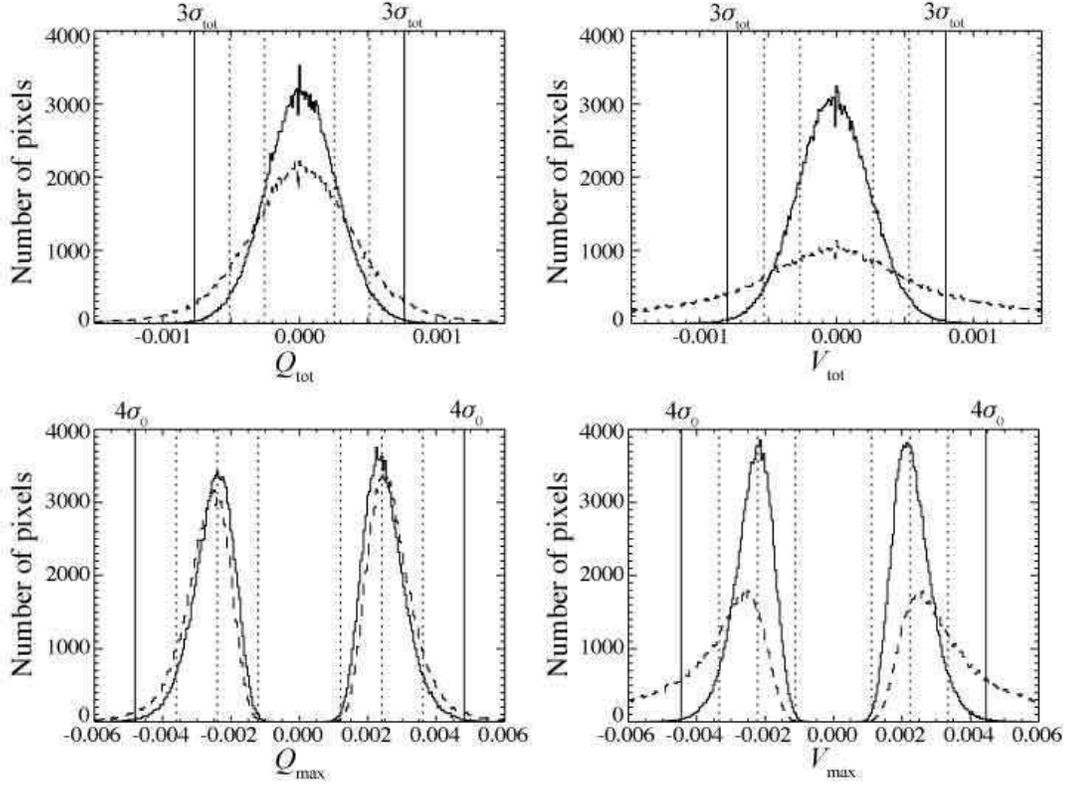}
\caption{Upper panels: Histograms (broken lines) of wavelength-integrated polarization signals ($Q_{\rm tot}$ and $V_{\rm tot}$)
and histograms (solid lines) of the noise ($Q'_{\rm tot}$ and $V'_{\rm tot}$).
Lower panels: Histograms (broken lines) of the peak polarization signals ($Q_{\rm max}$ and $V_{\rm max}$), and histograms
(solid lines) of the peak noise levels.}
\label{fig_noise}
\end{figure*}

The polarization signals in the internetwork region are typically small as compared with those in the 
network regions and active regions, and they are likely to be contaminated by the photon noise.
It is important to carefully choose pixels for analysis that are minimally affected by the noise.
First of all, we calculate the wavelength-integrated degree of polarization 
($Q_{\rm tot}$, $U_{\rm tot}$ and $V_{\rm tot}$) defined as,
\begin{equation}
Q_{\rm tot}\equiv \frac{\int_{\lambda_{b}}^{\lambda_{r}} Q(\lambda) d\lambda}{I_{c} \int_{\lambda_{b}}^{\lambda_{r}} d\lambda},
\label{eq:Qtot}
\end{equation}
\begin{equation}
U_{\rm tot}\equiv \frac{\int_{\lambda_{b}}^{\lambda_{r}} U(\lambda) d\lambda}{I_{c} \int_{\lambda_{b}}^{\lambda_{r}} d\lambda},
\label{eq:Utot}
\end{equation}
\begin{equation}
V_{\rm tot}\equiv \frac{\int_{\lambda_{b}}^{\lambda_{0}} V(\lambda) d\lambda}{I_{c} \int_{\lambda_{b}}^{\lambda_{0}} d\lambda} - \frac{\int_{\lambda_{0}}^{\lambda_{r}} V(\lambda) d\lambda}{I_{c} \int_{\lambda_{0}}^{\lambda_{r}} d\lambda},
\label{eq:Vtot}
\end{equation}
where $\lambda_{b}$ and $\lambda_{r}$ indicate the wavelength limits of the 
integration in blue and red wings of the \ion{Fe}{1} 630.25~nm line, respectively, 
and $\lambda_{0}$ represents the wavelength of the line center. 
The wavelength $\lambda_{0}$ corresponds to the minimum value of 
Stokes $I$ for $Q_{\rm tot}$ and $U_{\rm tot}$,
and to the zero-crossing wavelength of the Stokes $V$ profile for $V_{\rm tot}$.
Here we take $\lambda_{b}=\lambda_{0}-17.1$~pm, and $\lambda_{b}=\lambda_{0}+17.1$~pm.
In figure \ref{fig_noise}, the histograms of  
$Q_{\rm tot}$ and $V_{\rm tot}$ in our analyzed subfield (figure~\ref{fig_region}) are shown with dashed lines.
Since the histograms of $U_{\rm tot}$ are the same as those of 
$Q_{\rm tot}$, we do not show them.

To estimate the noise for $Q_{\rm tot}$, $U_{\rm tot}$, and $V_{\rm tot}$ 
we obtain the wavelength-integration
of the Stokes $Q$, $U$, and $V$ spectra ($Q'_{\rm tot}$, $U'_{\rm tot}$, and $V'_{\rm tot}$) 
in the wavelength range minimally affected by line polarization
(i.e., continuum range). 
The integrations are carried out over the same wavelength width of 36.4~pm. 
Then, we derive the noise level $\sigma_{\rm tot}$ which is the standard deviation 
of $Q'_{\rm tot}$, $U'_{\rm tot}$, and $V'_{\rm tot}$, and 
the value is estimated to be $2.6\times10^{-4}I_{c}$ for $Q_{\rm tot}$, 
$2.5\times10^{-4}I_{c}$ for $U_{\rm tot}$, 
and $2.7\times10^{-4}I_{c}$ for $V_{\rm tot}$.
Such wavelength-integrations reduce the noise level, and 
the sensitivity to weak magnetic field significantly increases.
The histograms of $Q'_{\rm tot}$, and $V'_{\rm tot}$ corresponding to the noise distributions are
shown with solid lines in the top panels of figure~\ref{fig_noise}.
Six vertical lines in each upper panel indicate 1, 2, and 3$\sigma_{\rm tot}$.
For our analysis, we choose the pixels with $Q_{\rm tot}$ or $U_{\rm tot}$ or $V_{\rm tot}$
above corresponding $3\sigma_{\rm tot}$ to avoid the pixels with false polarization signal (i.e., noise).

\citet{Viticchie2010AA} report that in the internetwork region, 
there are pixels which have heavily asymmetric Stokes profiles 
such as one lobe and three-lobe Stokes $V$ profiles in the \textit{Hinode} data.
One example is the footpoint regions of the horizontal magnetic fields in \citet{Ishikawa2010ApJ}.
Even if the maximum amplitudes of the Stokes profiles are large compared with the noise level $\sigma_{0}$
defined and estimated in Section \ref{obs},
wavelength-integrated polarization signals  are small, and may not surpass 
the threshold of $3\sigma_{\rm tot}$.
We include these pixels with high amplitude and low-integrated polarization signal in our analysis.
We measure the
maximum amplitudes of each Stokes profile of the \ion{Fe}{1} 630.25~nm line
($Q_{\rm max}$, $U_{\rm max}$ and $V_{\rm max}$)
to find the pixels with sufficient polarization signal induced by the magnetic fields. 
In the bottom two panels of figure~\ref{fig_noise}, 
the histograms of $Q_{\rm max}$ and $V_{\rm max}$ in our analyzed region
are shown with dashed lines, while the solid lines show
the histograms of maximum amplitudes of Stokes $Q$ and $V$ spectra
in the wavelength range 
without any polarization signal (i.e., continuum range).
These solid lines indicate the noise distributions of $Q_{\rm max}$ and $V_{\rm max}$. 
Eight vertical lines represent 1--4$\sigma_{0}$ for each Stokes profile.
Pixels with $Q_{\rm max}$ and $V_{\rm max}$ above 4$\sigma_{0}$
are essentially free from noise.
Since the histograms of $U_{\rm max}$ are the same as those of 
$Q_{\rm max}$, we do not show them.

In summary, we choose the valid pixels with
$Q_{\rm tot}$, $U_{\rm tot}$, or $V_{\rm tot}$ larger than $3\sigma_{\rm tot}$
\textit{or} with $Q_{\rm max}$, $U_{\rm max}$, or $V_{\rm max}$ larger than $4\sigma_{0}$. 
To further reduce the effect of noise, 
we choose patches that  contain at least 4 contiguous pixels, all of which meet the above criteria.
The size of 4 pixels corresponds approximately to
the instrument's spatial resolution.
According to this selection procedure, 10.1\% of the subfield has sufficient linear polarization signals 
(Stokes $Q$ and/or $U$),
and we refer to these pixels as horizontal field pixels (hereafter referred to as HF pixels).
Likewise, 43.8\% has sufficient circular polarization signal (Stokes $V$), and
these pixels are referred to as vertical field pixels (VF pixels).
Some pixels are identified as both the VF and the HF pixels.
In total, 46.9\% of the subfields are valid pixels in terms of above selection criteria.
This number (46.9\%) for Stokes $Q$, $U$ or $V$
is much larger than 35.5\% of \citet{Orozco2007ApJL} and 29\% of \citet{Viticchie2010AA}.
They investigated the same data set, 
but analyzed pixels only with the maximum amplitudes larger than 4.5 times their noise levels.

In the subfields, we identify 1085 HF patches and 1793 VF patches, respectively.
Among 1793 VF patches,
901 VF patches have positive polarity, and 892 VF patches have negative polarity. 
In figure~\ref{fig_mask}~(a), the VF patches with negative and positive polarities are shown in black
and white, respectively.
The red contours in the figure indicate the HF patches.

\subsection{Conversion of polarization signal to magnetic flux density}
\label{cal}
As a result of measurement noise, inversion of Stokes profiles for internetwork fields may not converge, thus compromising the uniqueness and accuracy of the solution. 
Furthermore, the asymmetric profiles often seen in the internetwork region make the proper inversion more difficult.
Indeed, Milne-Eddington inversion, which has been widely used for the analysis of the SOT/SP data, cannot deal 
with the heavily asymmetric Stokes profiles.
More powerful inversion techniques such as 
SIR \citep[Stokes Inversion based on Response function,][]{RuizCobo1992ApJ} 
and MISMA \citep[MIcro-Structured Magnetic Atmosphere,][]{SanchezAlmeida1997ApJ}
are useful to interpret the Stokes profiles with large asymmetries.
However, in these techniques 
the number of fitting parameters increases,
and we may potentially face the severe problem of the uniqueness of the solution.

Therefore, in this paper, we do not use the inversion technique to retrieve the properties of
the internetwork magnetic fields. 
Instead, we follow a procedure to convert the wavelength-integration of the polarization signal into
the longitudinal and transverse magnetic flux density.
Following previous studies \citep[e.g.,][]{Lites2008ApJ}, 
we refer to these two measurements as ``apparent flux density", $B^{L}_{\rm app}$ and
$B^{T}_{\rm app}$,
and use the units of Mx cm$^{-2}$ for them, while the unit of gauss is used 
for intrinsic field strength.
This procedure does not allow us to resolve fine magnetic structure within the resolution element, 
which is usually represented by the magnetic filling factor $f$.
$B^{L}_{\rm app}$ and $B^{T}_{\rm app}$ are
average field strengths over the resolution element,
and are also interpreted as the magnetic field strength assuming 
that $f=1$.

The wavelength-integrated circular polarization $V_{\rm tot}$ defined by equation (\ref{eq:Vtot})
is converted to $B_{\rm app}^{L}$.
We also define the net linear polarization $L_{\rm tot}$ which will be converted to $B_{\rm app}^{T}$;
 \begin{equation}
L_{\rm tot}\equiv \frac{\int_{\lambda_{b}}^{\lambda_{r}} \sqrt{Q(\lambda)^2+U(\lambda)^2} d\lambda}{I_{c} \int_{\lambda_{b}}^{\lambda_{r}} d\lambda}.
\label{eq:Ltot}
\end{equation}
Here $\lambda_{b}$ and  $\lambda_{r}$ are defined in Section \ref{pix}.

\citet{Lites1999ApJ} and \citet{Jin2009ApJ} employed the inversion results for pixels with stronger polarization signal primarily from network regions in the quiet Sun to calibrate the relationship of
wavelength-integrated polarization signals to the apparent flux densities.
However, the intrinsic field strengths of the network fields are typically stronger than $\sim$1~kG, and
the calibration curve derived from the network fields may not be applied to the internetwork regions which harbor hG magnetic fields \citep[e.g.,][]{Orozco2007PASJ,Orozco2007ApJL}. 

\begin{figure*}
%\epsscale{1.5}
\epsscale{1.0}
\plotone{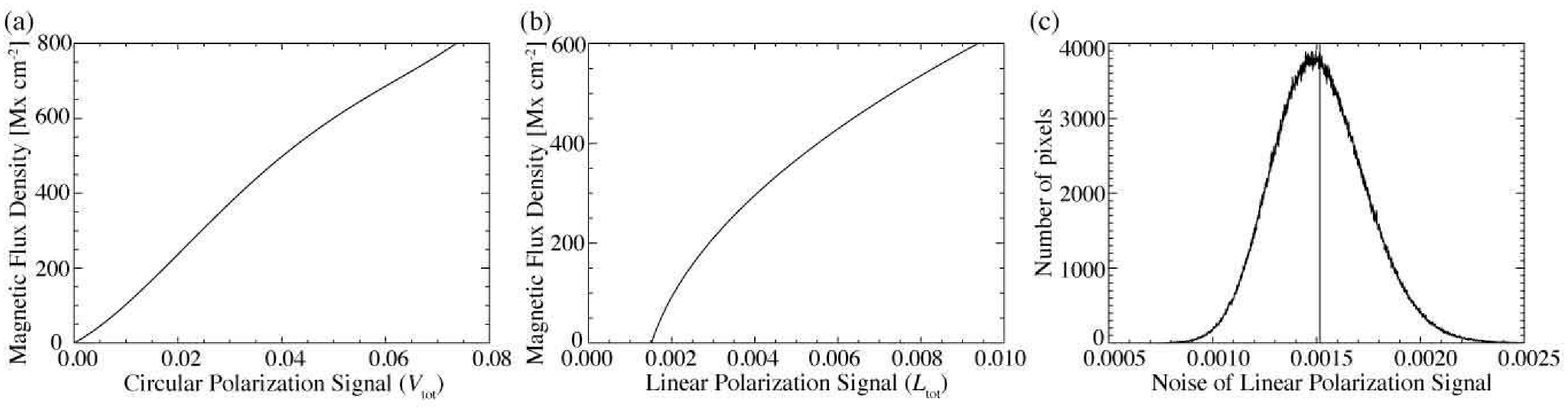}
\caption{Calibration curves derived from the synthetic Stokes profiles 
based on Milne-Eddington atmosphere for weak fields
 to convert the wavelength-integrated polarization signals 
 ($V_{\rm tot}$ and $L_{\rm tot}$) to magnetic flux density 
 for longitudinal magnetic components~(a) and 
transverse components~(b).
(c) Noise distributions of $L_{\rm tot}$.}
\label{fig_calibration}
\end{figure*}

The approach to determine the longitudinal and transverse magnetic flux density
$B^{L}_{\rm app}$ and $B^{T}_{\rm app}$ is the following \citep[e.g.,][]{Lites2008ApJ}. 
We calculate the synthetic Stokes profiles based on the Milne-Eddington atmosphere.
For the calculation, 
the intrinsic field strength $B$ is set to be smaller than 1~kG, the stray light fraction, 
$\alpha$ ($\alpha=1-f$) to be smaller than 0.6, 
and the absolute value of Doppler velocity $v_{\rm LOS}$ to be less than 2~km~s$^{-1}$.
Here we choose the field strengths, with which the weak field approximation is valid. 
We employ relatively large filling factor ($f>0.4$) 
since the small filling factor with weak field strength does not produce enough Stokes signals.
Many synthetic profiles were generated for $0<B<1$~kG, $0.4<f<1.0$, $-2.0<v_{\rm LOS}<2.0$~km~s$^{-1}$ and $0<\gamma<180^{\circ}$,
where $\gamma$ is the inclination
angle with respect to the line-of-sight (LOS) direction.
Following equations (\ref{eq:Vtot}) and (\ref{eq:Ltot}), we calculate $V_{\rm tot}$ and $L_{\rm tot}$ 
from the synthetic profiles.
The longitudinal and transverse flux density $B^{L}_{\rm app}$ and $B^{T}_{\rm app}$  are obtained by 
\begin{equation}
B^{L}_{\rm app}=fB\cos\gamma,
\label{eq:BL}
\end{equation} 
and
\begin{equation}
B^{T}_{\rm app}=\sqrt{f}B\sin\gamma.
\label{eq:BT}
\end{equation} 
The wavelength-integrated degree of polarization $V_{\rm tot}$ and $L_{\rm tot}$ 
are compared with $B^{L}_{\rm app}$ and $B^{T}_{\rm app}$  
to derive the calibration curves optimized for the internetwork fields (Figure \ref{fig_calibration}).
Since the observing region is located at disk center, $B^{T}_{\rm app}$ and  $B^{L}_{\rm app}$ 
correspond to the horizontal and vertical magnetic flux density, respectively.

The measurement noise $\sigma_{0}$ at each wavelength position is not canceled out, and 
coherently contributes to the 
integration over the wavelength for $L_{\rm tot}$ (equation (\ref{eq:Ltot})).
Figure~\ref{fig_calibration}~(c) shows the noise distribution of the net linear polarization signal 
$L_{\rm tot}$, which is  
derived with equation~(\ref{eq:Ltot}) but with the integration carried out over 
the continuum range.
We set $B_{\rm app}^{T}=0$ for  
the mean value $L_{\rm tot}=1.51\times10^{-4}I_{c}$,
which is indicated by the solid line in figure~\ref{fig_calibration}~(c).
In general, the distribution function $p(l)$ with $l=\sqrt{q^{2}+u^{2}}$, 
where $q$ and $u$ have the Gaussian distributions with the same standard deviation $\sigma$ 
($p(q)=\frac{1}{\sqrt{2\pi}\sigma}\exp(-\frac{q^2}{2\sigma^2})$ and
$p(u)=\frac{1}{\sqrt{2\pi}\sigma}\exp(-\frac{u^2}{2\sigma^2})$) is written as
$p(l)=\frac{l}{\sigma^2}\exp(-\frac{l^2}{2\sigma^2})$.
Thus, the mean value of $l$ is slightly deviated from the peak of the distribution as shown in  figure~\ref{fig_calibration}~(c).
Note that this calibration will underestimate the magnetic flux density
where the intrinsic field strengths surpass kG strengths, as it does in network regions.

\begin{figure*}
%\epsscale{1.5}
\epsscale{1.0}
\plotone{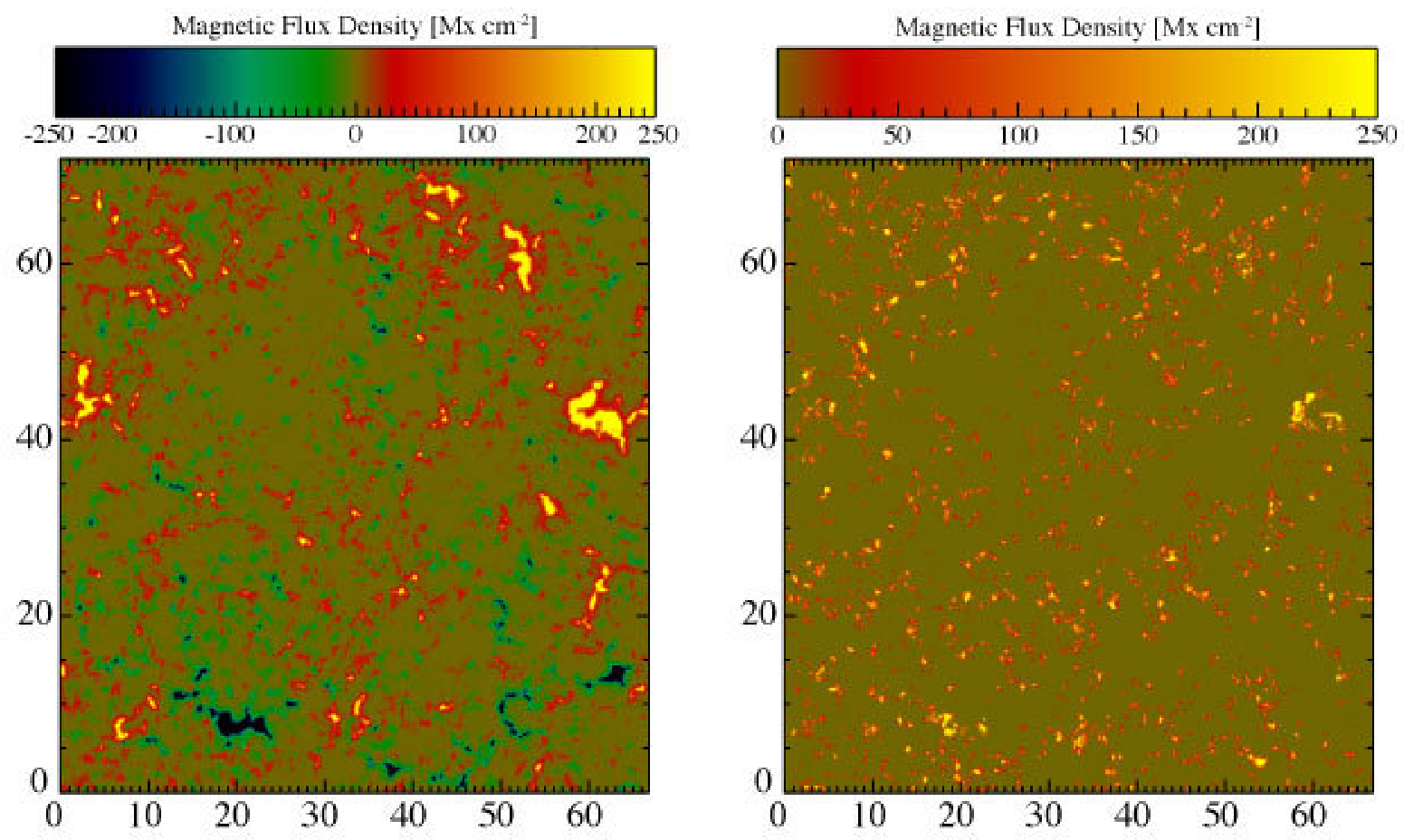}
\caption{Longitudinal ($left$) and transverse ($right$) magnetic flux density, 
$B_{\rm app}^{L}$ and $B_{\rm app}^{T}$,
for the  a $67\farcs5\times72\arcsec$ subfield shown in figure~\ref{fig_region}.
Color scales for both panels are the same and are saturated at 250~Mx~cm$^{-2}$.
We set  $B_{\rm app}^{L}=0$ in pixels which are not identified as the VF pixels, 
and set $B_{\rm app}^{T}=0$ in pixels which are not identified as the HF pixels (Section \ref{pix}).}
\label{fig_field}
\end{figure*}

Figure~\ref{fig_field} shows the spatial distributions of longitudinal (vertical) and transverse (horizontal)
magnetic flux density, $B_{\rm app}^{L}$ and $B_{\rm app}^{T}$
having been obtained through the procedure described above.
The scales for both components are saturated at 250~Mx~cm$^{-2}$ in order to compare 
the two maps in the internetwork region. 

We identify the pixels with $V_{\rm tot}>0.5\%$ and the area $\ge$90~pix$^{2}$
as network pixels, 
which are marked with yellow contours in figure~\ref{fig_mask} (b).
HF patches that overlap with the network pixels 
are indicated with green contours in figure~\ref{fig_mask} (b). 
These HF patches may be the part of  
the canopies of the network fields.
We do not include these HF patches in the subsequent analysis.
The red contours in figure~\ref{fig_mask}~(b) indicate 
the horizontal magnetic fields (HF patches) in the internetwork regions.
%%%%%%%%%===============

\section{Close association of horizontal magnetic patches with vertical magnetic patches}
\label{association}
\begin{figure*}
%\epsscale{1.5}
\epsscale{0.8}
\plotone{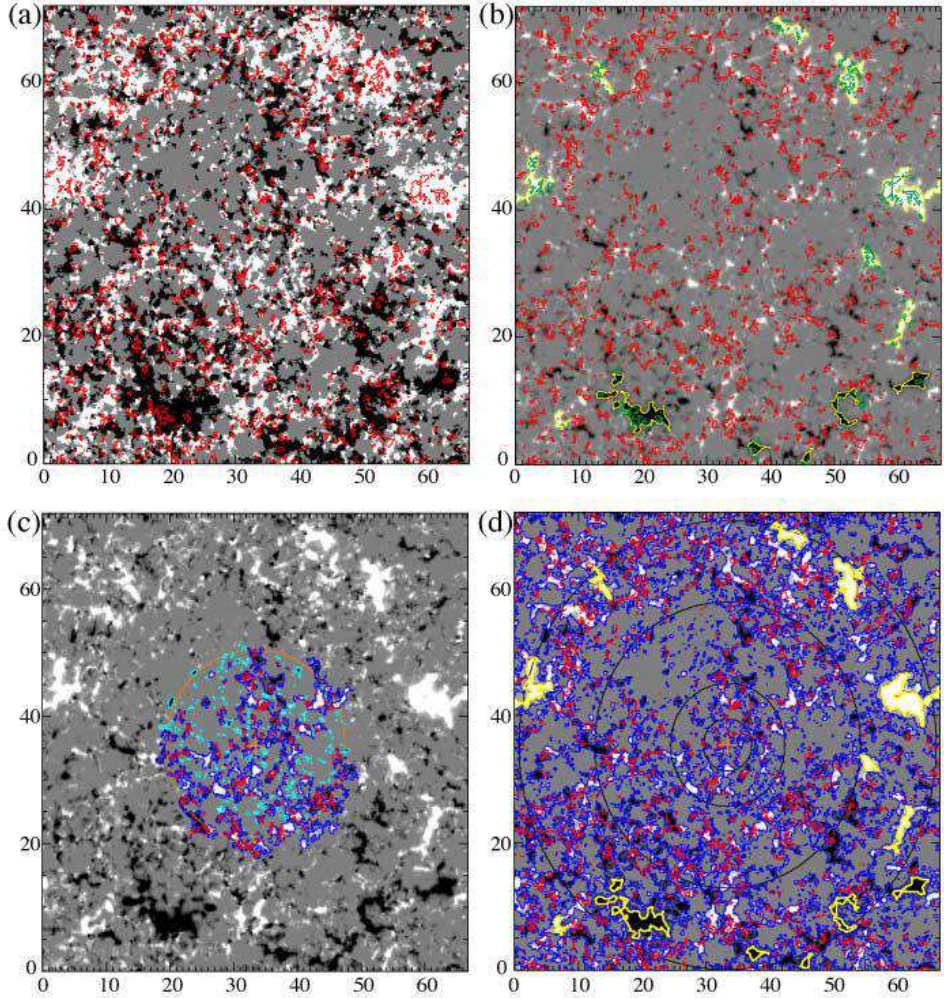}
\caption{(a) VF patches with negative and positive polarities
are shown in black and white. 
The red contours represent HF patches.
(b) Circular polarization map ($V_{\rm tot}$) saturated at $V_{\rm tot}=\pm0.01$.
The red contours are the same as the panel (a).
Yellow contours indicate the network pixels.
Green and red contours indicate the HF patches
in the network region, and the HF patches in the internetwork region.
In panels (c) and (d), 
circular polarization map ($V_{\rm tot}$) saturated at $V_{\rm tot}=\pm0.005$ is shown and
the red contours indicate the HF patches. 
(c) The cross mark indicates an approximate center position of a supergranular cell. 
The region enclosed by the orange circle is used for the estimation
of magnetic flux in the internetwork region.
Dark blue contours indicate the VF patches associated with the HF patches, 
while light blue ones indicate the VF patches which do not overlap with the HF patches.
(d) The blue contours indicate the VF regions, while
yellow contours indicate the network magnetic fields.}
\label{fig_mask}
\end{figure*}

In this section, we examine the spatial relationship 
between the horizontal and vertical magnetic fields.
Figure~\ref{fig_mask}~(a) and figure~\ref{fig_partial}~(a) indicate that both horizontal and vertical magnetic
components together form the cellular structures with a spatial scale of $5\arcsec-10\arcsec$ \citep{Lites2008ApJ}.
We also notice that the HF patches overlap with the VF patches.

As mentioned in section~\ref{pix}, 
the total number of VF pixels in total is 4.3 times as large as that of the HF pixels,
and 69.4\% of the HF pixels overlap the VF pixels.
Alternatively, 48.7\% of the total number of the VF patches has cospatial HF patches.
The size of the VF patches is apparently larger than that of the HF patches (figure~\ref{fig_mask}~(a)).
This would be because the sensitivity of circular polarization to detect VF patches 
is much higher than that of linear polarization to detect HF patches.
Thus we detect only the core portion of the HF patches.
Even though many HF patches overlap with the VF patches,
they are not contained in the VF patches, and
tend to be located 
around the edge of the VF patches (figure~\ref{fig_mask}~(a)). 
We quantitatively examine where 
the HF patches are located with respect to the VF patches by means of the procedure used in \citet{Ishikawa2007AA}.

\begin{figure*}
%\epsscale{1.5}
\epsscale{1.0}
\plotone{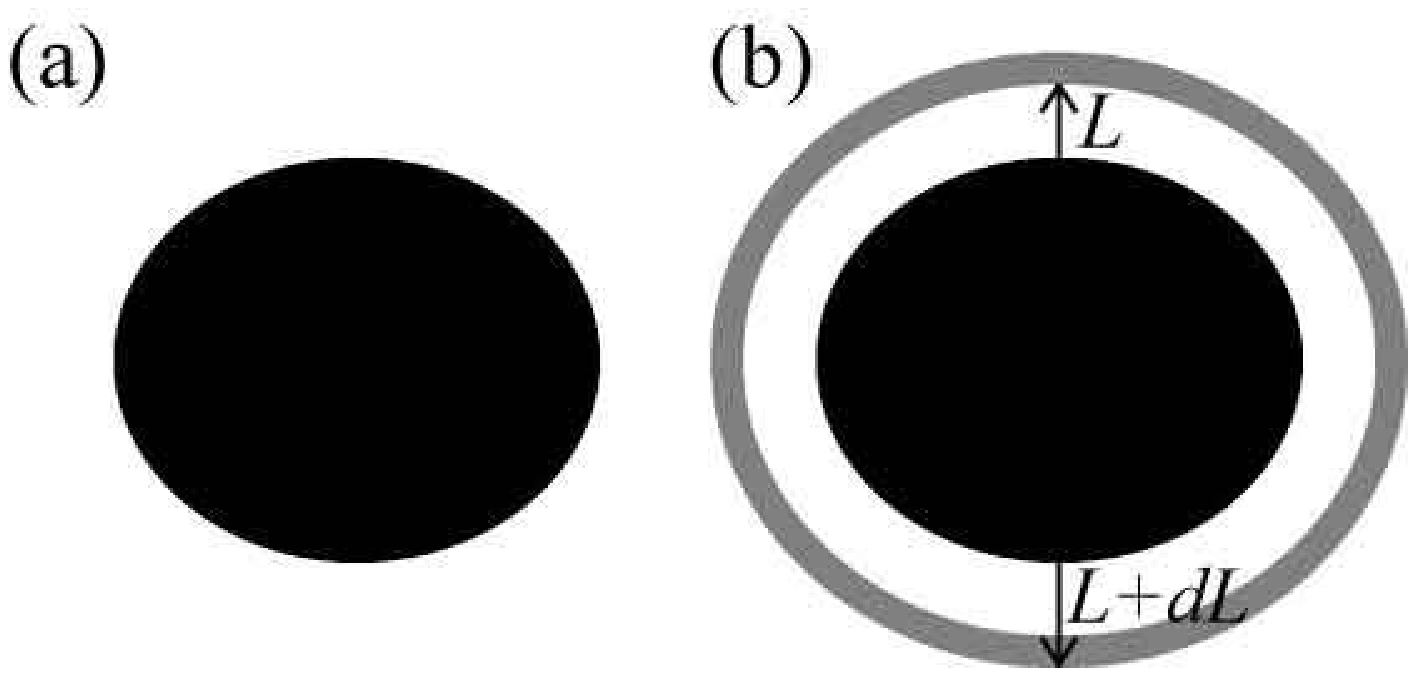}
\caption{(a) Schematic diagram of a VF region.
(b) Schematic diagram to obtain $N_{i}(L)$ ($i=\rm all, \rm HF$)
with the outward distance $L$ from the edge of the VF region.
$L$ is  the outward distance from the edge of the VF region.
See text for details.}
\label{fig_schematic}
\end{figure*}

The area density $D(L)$ of the HF patches in the internetwork region at
the distance $L$ ($L<0$ inward, and $L>0$ outward) 
from the borders of the VF patches with width $dL$ (figure \ref{fig_schematic}) is given by
\begin{equation}
D(L)=\frac{N_{\rm HF}(L)}{N_{\rm all}(L)},
\label{eq:D}
\end{equation}
where $N_{\rm all}(L)$ is the total number of pixels that form the VF patches
between $L$ and $L+dL$, and 
$N_{\rm HF}(L)$ is the total number of pixels
occupied by the HF patches located in the distance between $L$ and $L+dL$.
Here we take $dL=1$~pixel (0\farcs16).
Note that $N_{\rm all}(L)$ and $N_{\rm HF}(L)$ are not obtained for a single VF patch, 
but for all the VF patches.

The procedure to obtain $N_{i}(L)$ ($i=\rm all, \rm HF$) with the outward direction ($L>0$) is as follows.
Let us consider the case with one VF patch shown in black in figure~\ref{fig_schematic}~(a).
$N_{\rm all}(L)$ is the total number of pixels in the gray area of figure~\ref{fig_schematic}~(b)
and $N_{\rm all}(L)$ is the total number of pixels occupied by the HF patches in the same gray area.
First, we obtain 
the total area of the black and white region (figure~\ref{fig_schematic}~(b)) $S_{\rm all}(L)$, 
and the total area of the HF patches in the black and white region $S_{\rm HF}(L)$.
Second, we obtain $S_{\rm all}(L+dL)$, which is the total area of the black, white, and the gray region (figure~\ref{fig_schematic}~(b)),
and  $S_{\rm HF}(L+dL)$, which is the total area occupied by HFs in the same black, white, and gray region.
Finally, the areas $N_{i}(L)$ with $i=\rm all, \rm HF$ are given by $N_{i}(L)=S_{i}(L+dL)-S_{i}(L)$.
Similarly we obtain 
$N_{i}(L)$ with the inward direction ($L<0$).

\begin{figure*}
%\epsscale{1.5}
\epsscale{1.0}
\plotone{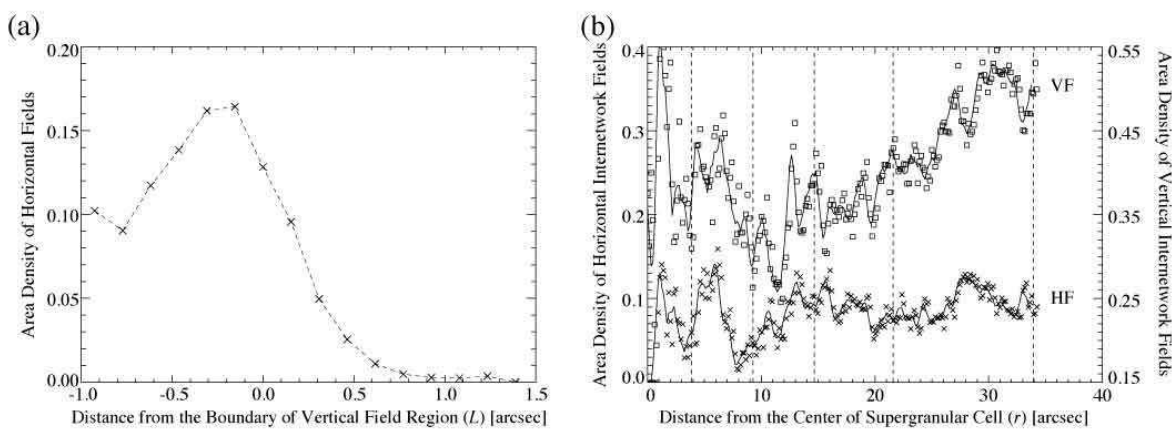}
\caption{(a) Area density $D(L)$ [eq.(\ref{eq:D})] of the HF patches in the internetwork region as a function of the distance from
the boundary of the VF patches.
The distance of $L=0$ corresponds to the edge of the VF patches, and
the negative and positive distances mean inward and outward directions, respectively.
(b) Area density of the VF and HF patches in the internetwork region as a function of the distance from
the center position of a supergranular cell.
Solid lines indicate the running mean with 4~pixels ($\sim0\farcs6$).}
\label{fig_plot}
\end{figure*}

Figure~\ref{fig_plot}~(a) shows the area density $D(L)$ of the HFs derived with this procedure.
There are more HFs around the boundary
of the VF pixels;
the area density of the HFs reaches its maximum point at the inward distance of
 $L=-0\farcs{16}$ (1~pixel)
and rapidly decrease over $L\sim$0\farcs6.
This clearly shows a remarkable association of the smaller HF patches with the larger VF patches at their borders.

We detect 1085 HF patches in the internetwork region, and 53\% of them have circular polarization signals 
($V_{\rm tot}$) of both polarities inside the HF patches and their vicinity (within outward distance of one pixel $\sim$$0\farcs16$).
The HF patches that have one polarity comprise 
another 43\% of the HF patches.
The rest of the HF patches (4\%) do not have any circular polarization ($V_{\rm tot}$).
A similar result is reported by \citet{Danilovic2010ApJL} using the Sunrise/IMax instrument.
Horizontal magnetic fields without accompanying vertical components are reported by \citet{Jin2009ApJ}.
Such HFs were found by \citet{Jin2009ApJ} for $\sim$20\% of the detectable events, 
a percentage that is much higher than the one from our own analysis (4\%).
In our analysis, we chose the pixels for which polarization signals are as weak as possible,
yet still have minimal influence from noise.
This careful treatment should have enabled us to find a larger number of HF patches without vertical components in the internetwork region if the results of \citet{Jin2009ApJ} were correct.
In Figure~\ref{fig_partial}~(a), HF patches with $V_{\rm tot}$ of both polarities
(bipolar HF patches) are shown with red contours,
while HF patches with $V_{\rm tot}$ of one polarity (unipolar HF patches) are shown with green contours.
The HF patches without $V_{\rm tot}$ (isolated HF patches) are indicated with blue contours.

\begin{table*}
\caption{The properties of HF patches. 
The values in brackets indicate the total number of analyzed HF patches.
SD represents standard deviation.}
\label{table1}
\begin{center}
\begin{tabular}{ccccccccc}
\hline & \multicolumn{2}{c}{Bipolar (574)} &  \multicolumn{2}{c}{Unipolar (467)} &  \multicolumn{2}{c}{Isolated (43)} &  \multicolumn{2}{c}{All (1084)} \\ 
& mean & SD & mean & SD & mean & SD & mean & SD\\ \hline \hline
max $B_{app}^{\rm L}$ (Mx~cm$^{-2}$) & 39 & 52 & 47 & 69 & \multicolumn{2}{c}{$-$} & 41 & 58\\ \hline
max $B_{app}^{\rm T}$ (Mx~cm$^{-2}$) & 157 & 60 & 112 & 38 & 91 & 24 & 135 & 56\\ \hline
Area (arcsec$^{2}$) & 0.58 & 0.72 & 0.19 & 0.17 & 0.12 & 0.04 & 0.39 & 0.57\\ \hline
\end{tabular}
\end{center}
\end{table*}

\section{Properties of horizontal magnetic patches}
\label{property}
\begin{figure*}
%\epsscale{1.5}
\epsscale{1.0}
\plotone{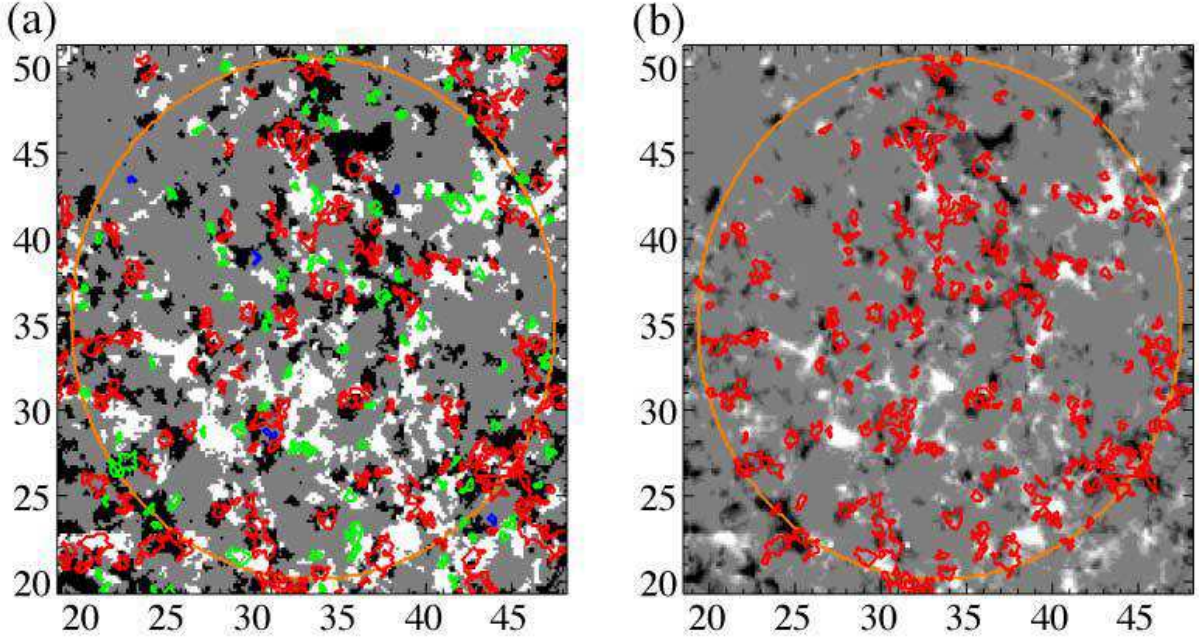}
\caption{Enlargement of the center part of the internetwork region.
The orange circle is the same as that in panel (c) of Figure~\ref{fig_mask}.
(a) VF patches with negative and positive polarities are shown in black and white. 
Bipolar HF patches are shown with red contours, while unipolar ones are shown with green contours.
HF patches without vertical magnetic components are shown with blue contours.
(b) The circular polarization map ($V_{\rm tot}$) saturated at $V_{\rm tot}=\pm0.005$ is shown and
the red contours indicate the HF patches.}
\label{fig_partial}
\end{figure*}

\begin{figure*}
%\epsscale{1.5}
\epsscale{1.0}
\plotone{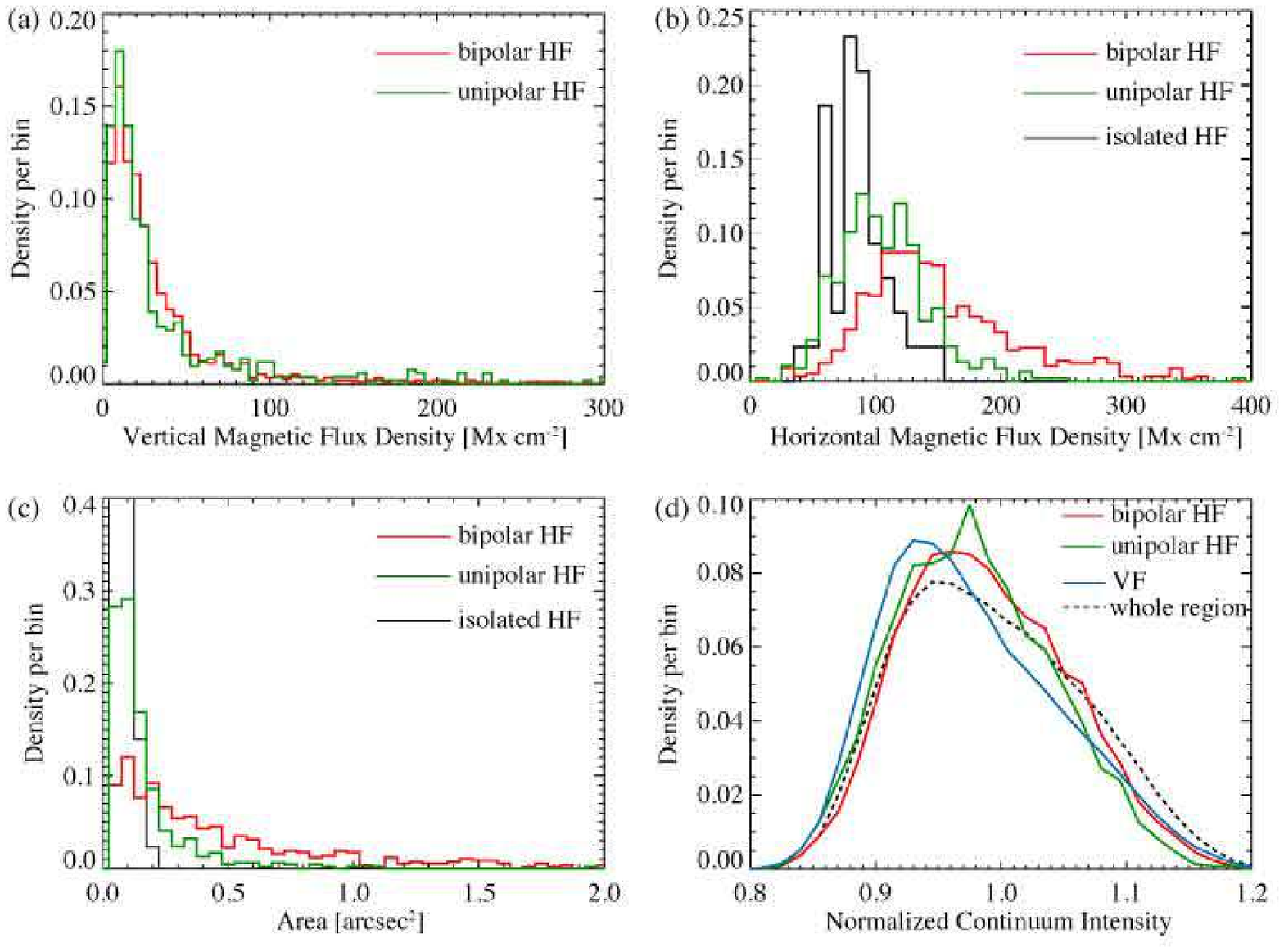}
\caption{(a) Histograms of maximum vertical magnetic flux density in the HF patches.
(b) Histograms of maximum horizontal magnetic flux density in the HF patches.
(c) Histograms of the HF patch area.
(d) Histograms of the normalized continuum intensities in the HF region.}
\label{fig_hist}
\end{figure*}

We investigate whether the properties of the HF patches such as their magnetic flux density, 
their size, and their locations
depend on the presence of the vertical magnetic components. 
Figure~\ref{fig_hist}~(a) indicates the normalized histograms of the maximum unsigned vertical magnetic 
flux density inside the HF patch. For bipolar HF patches (red lines), 
the histogram includes the maximum value of the positive magnetic flux density
and that of the unsigned negative magnetic flux density.
In the weak end of vertical magnetic flux  density ($<\sim$100 Mx cm$^{-2}$),
there is no significant difference between unipolar HF events (green lines) and bipolar ones (red lines).
However, more unipolar events harbor relatively strong 
vertical magnetic flux density ($>100$ Mx cm$^{-2}$).
These events would represent the canopy structures
of the vertical magnetic elements, and
one example is seen in the unipolar HF patch at (27\arcsec, 46\arcsec)
in figure~\ref{fig_partial}~(a).

Figure~\ref{fig_hist}~(b) shows the histogram of the maximum horizontal
magnetic flux density in the bipolar (red lines),
the unipolar (green), and the isolated HF patches (black).
Here the isolated HF patches refer to those without detectable circular polarization signal.
The maximum horizontal magnetic flux density of the bipolar HF patches ranges
up to 350~Mx~cm$^{-2}$ with an average value of 
$\sim$157~Mx~cm$^{-2}$, while the maximum horizontal magnetic flux density
of the unipolar HF patches ranges to 200~Mx~cm$^{-2}$
with an average value of  $\sim$112~Mx~cm$^{-2}$ (see also table~\ref{table1}).
On average, the maximum horizontal flux density of the bipolar HF patches
is larger than that of the unipolar HF patches.
The flux density of the isolated HF patches is essentially below 150~Mx~cm$^{-2}$
with an average value of 91~Mx~cm$^{-2}$,
and is much weaker than that of HF patches associated with a vertical component.

The histograms of the area of these patches are shown in figure~\ref{fig_hist}~(c).
The area of the bipolar HF patches distributes over the wide range, 
while the area of the unipolar HF patches is essentially limited to 1.0 arcsec$^{2}$
and 74\% of the unipolar HF patches are smaller than 0.2 arcsec$^{2}$.
The area of all the isolated events is smaller than 0.2 arcsec$^{2}$.
The number of HF patches generally increases with decreasing size, 
and about 20~\% of the HF patches have a size smaller than 0.1~arcsec$^{2}$ (4~pix), 
which is comparable to the SOT/SP spatial resolution
($0\farcs32\times0\farcs32$).
This indicates that there are a number of 
HF patches that are not resolved with SOT/SP.

Figure~\ref{fig_hist}~(d) shows the normalized continuum intensity distributions
for bipolar and unipolar HF patches, 
and the VF patches (blue line) in the internetwork region.
It appears that the continuum intensity distributions for both bipolar and unipolar HF patches 
have peaks slightly brighter than that for the whole subfield (dashed line),
and are higher than the intensity distribution of the whole subfield
at intermediate intensities ($\sim$0.95 -- $\sim$1.05).
This is consistent with the observations that HF patches are statistically 
located in the outer boundary of the granules 
\citep{Lites2008ApJ,Centeno2007ApJL,Ishikawa2008AA,Ishikawa2009ASPC,Danilovic2010ApJL}.
We do not display the continuum intensity distribution of the isolated HF patches, 
which is noisy due to the small number of pixels.
It appears that (1) the continuum intensity distribution of VF patches has a peak darker than that of the whole subfield,
and is higher at the lower intensities ($\sim$0.85 -- $\sim0.95$)
and that (2) the continuum intensities of VF patches are
darker than those of HF patches, and VF patches could be located closer to the 
intergranular lane.

The unipolar HF events 
have lower horizontal magnetic flux densities, and a smaller area than the bipolar events.
Both bipolar and unipolar events tend to be located in the periphery of granules.
As to the isolated HF patches, the horizontal magnetic flux densities
are low and their areas are small compared with the bipolar and unipolar HF patches.  
Although some unipolar HF patches correspond to the canopy structures 
of the vertical magnetic fields, 
it is conceivable that the some unipolar and isolated HF patches are
essentially the same phenomena as the bipolar patches;
they have the loop structure with bipolar footpoints.
Since the sensitivity to horizontal field is low compared with sensitivity to the vertical field,
only cores of the HF patches attached to one of the footpoints are detected as unipolar HF patches.
This would result in the smaller size of the unipolar HF patches.
The candidate for such HF patches is found at (25\arcsec, 43\arcsec)
in figure~\ref{fig_partial}~(a).
When the HF patches have a size closer to the resolution element of the telescope, 
their circular polarization signals with opposite polarities 
corresponding to their footpoints start to be canceled out.
As a result, the residual Stokes $V$ signal becomes weak, and 
it is possible that only circular polarization signals with one dominant polarity are left, 
or all the Stokes $V$ signals are completely canceled out.
%%%%%===========================

\section{Estimation of horizontal and vertical magnetic fluxes}
\subsection{Global magnetic flux balance of VF and HF patches}
\label{globalflux}
In this section, we compare the total magnetic fluxes to clarify which component 
dominates the internetwork region, vertical or horizontal.
We derive  and compare the total vertical and horizontal magnetic flux 
in the pure internetwork region. 
In figure~\ref{fig_mask}~(c), the cross mark indicates the center position of a supergranular cell,
and we define the region within a distance of $15\arcsec$ from the center (orange circle)
as the pure internetwork region. 
Some vertical patches crossing over the orange line are also included in the analysis.
We consider the HF patches that are located inside the orange circle 
and overlap with the VF patches (dark blue contours in figure~\ref{fig_mask}~(c)).
The VF patches without HFs are shown with the light
blue contours in figure~\ref{fig_mask}~(c).

The total unsigned vertical magnetic flux within the orange circle is
given by $\sum |B\cos\gamma f| ds dl=\sum |B_{\rm app}^{L}|dsdl$, 
where $B$ is the intrinsic magnetic field strength,
$f$ the filling factor,
$ds$ the scanning step size of $0\farcs15$,
and $dl$ is the pixel scale along the slit of 0\farcs16.
The total flux is estimated to be 3.0$\times$10$^{19}$~Mx~cm$^{-2}$ and 
the average flux density is 8.3~Mx~cm$^{-2}$.
If we consider only the VF patches that have the HF patches, 
total vertical magnetic flux is estimated to be 2.7$\times$10$^{19}$~Mx~cm$^{-2}$, 
and the average flux density is  7.5~Mx~cm$^{-2}$.
We reach the remarkable conclusion that most ($\sim$91\%) 
of the vertical magnetic flux are contributed by the VF patches 
accompanied by the HF patches.
The mean flux density obtained here is comparable to the one obtained by
\citet[][9.5~Mx~cm$^{-2}$]{Orozco2007ApJL} and 
\citet[][11.7~Mx~cm$^{-2}$]{Lites2008ApJ}.

Following \citet{Lites2008ApJ}, we also derive the spatial average horizontal 
flux density
defined by $\sum B_{\rm app}^{T}dsdl/\sum dS$, where $\sum dS$ is the total area
enclosed by the orange circle. 
The average horizontal flux density is 7.1~Mx~cm$^{-2}$ 
which is comparable to the vertical one
derived above and much smaller than the one obtained by \citet[60~Mx~cm$^{-2}$]{Lites2008ApJ}.

Unlike the VF case, the horizontal magnetic flux cannot be estimated by 
$B_{\rm app}^{T}$ times the pixel area,
since the magnetic fields are directed perpendicular to the LOS.
If all the identified horizontal magnetic patches have a flux tube structure \citep{Ishikawa2010ApJ},
the cross section $wdz$ of the individual flux tubes instead of the apparent size should be used for the estimation of the total magnetic
flux, 
where $w$ is the lateral size of the flux tube and $dz$ the thickness along the LOS.
Assuming that the flux tube is spatially resolved 
and the magnetic filling factor $f$ is the fraction occupied by the magnetized plasma along the LOS
in the line-forming layer,
the total magnetic flux of an individual event is estimated by $\overline{B\sin\gamma} w dz\sim\overline{B\sin\gamma}w f L$,
where $\overline{B\sin\gamma}$ is the average horizontal intrinsic field strength in the HF patch
and $L$ is the length of the line forming region ($\sim$300~km).
Using equation (\ref{eq:BT}), the total magnetic flux is 
written as $\overline{B_{\rm app}^{T}}\sqrt{f}wL$,
where $\overline{B_{\rm app}^{T}}$ is the average transverse (horizontal) magnetic flux density 
in the HF patch.
Since $f$ is not derived in our analysis, we use the average filling factor $f\sim0.15$ of the HFs in the quiet Sun
obtained with Milne-Eddington inversion \citep{Ishikawa2009AA}.
We fit a spatial ellipse to all the HF patches and use the minor axes for
the lateral size of the flux tube $w$.
The total magnetic flux of the HF patches within the orange circle taking into 
account their cross sections is estimated to be 4.7$\times$10$^{18}$~Mx.
This value is smaller than
the total unsigned vertical magnetic flux (3.0$\times$10$^{19}$~Mx)
by a factor of 6. 
Since the positive and negative magnetic flux are grossly balanced inside the orange circle,
total magnetic flux for the VF patches will be three times larger than that for the HF patches.
This value could be larger or smaller,
since we assume the nominal filling factor of 0.15 
and the thickness of the line-forming layer of 300~km, and
roughly estimate the cross section of the flux tubes.

\subsection{Local magnetic flux balance of VF and HF patches}
\begin{figure*}
%\epsscale{1.5}
\epsscale{1.0}
\plotone{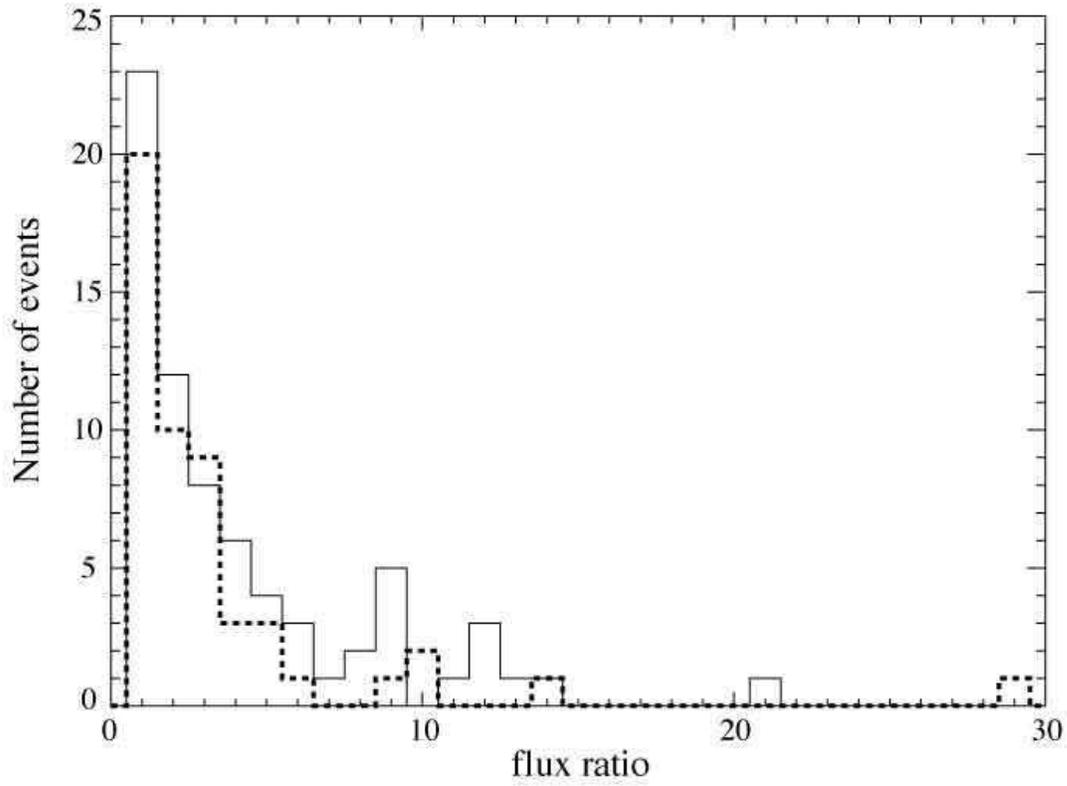}
\caption{Histograms of magnetic flux ratio when the VF flux is larger than the HF flux (solid line) and the HF flux is larger than the VF flux (dashed line).}
\label{fig_fluxratio}
\end{figure*}

Given the fact that the HF patches are closely associated with the VF patches (Section \ref{association}),
we quantitatively examine the magnetic flux of the VF patches associated with each HF patch.
We estimate the magnetic flux of the HF patches by considering the cross section of the flux tubes
as was done in Section \ref{globalflux}.
We identify the overlapping HF and VF patches, and 
plot the ratio of HF  and VF fluxes in figure~\ref{fig_fluxratio}.
There are cases where an individual HF patch overlaps with several VF patches, and vice versa.
If the VF magnetic flux is larger than the HF flux, we take the ratio of the VF flux to the HF flux (solid line in figure~\ref{fig_fluxratio}).
If the HF flux is larger than the VF flux, 
we take the ratio of the HF flux to the VF flux (dashed line in figure~\ref{fig_fluxratio}). 
The histograms in figure~\ref{fig_fluxratio} have the peak at 1, indicating that
the magnetic fluxes of associated HF and VF patches are well balanced.

\section{Intrinsic Magnetic Field Strength}
\begin{figure*}
%\epsscale{2.0}
\epsscale{1.0}
\plotone{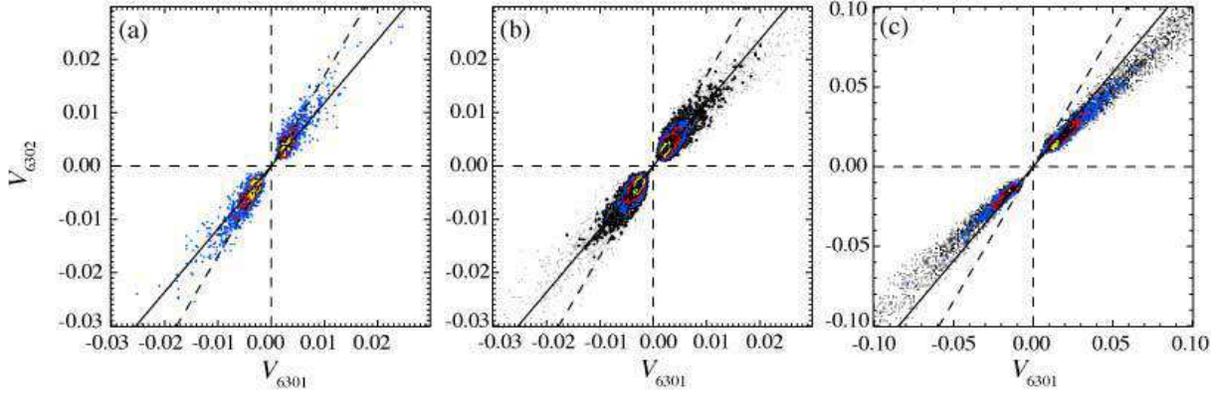}
\caption{Scatter plot of the maximum blue lobe Stokes $V$ amplitudes derived for the 
630.15~nm line vs. the corresponding Stokes $V$ amplitudes for the 630.25~nm line. 
The dashed line with slope $s=1.69$ represents the regression relation
that would be expected for  weak fields.
Solid lines have the slope of $0.7s$.
(a) the VF pixels that coincide in position  with the HF pixels inside the orange circle
in figure~\ref{fig_mask}~(c). 
(b) the VF pixels not overlapping with HF pixels inside the orange circle
in figure~\ref{fig_mask}~(c). 
(c) network fields shown with yellow contours in figure~\ref{fig_mask}~(d).
The contours indicate the density of data points.
The yellow (red, blue and black) contours indicate the area with the density higher than
0.3\% (0.1\%, 0.05\% and 0.01\%) of the total number of data points.}
\label{fig_ratio}
\end{figure*}

In this section, we examine the Stokes $V$ amplitude line ratio $V_{6302}/V_{6301}$
to infer intrinsic field strength \citep{Stenflo2010AA}.
Figure~\ref{fig_ratio} shows the scatter plots of the maximum Stokes $V$ amplitudes
in  the blue lobe of  \ion{Fe}{1} 630.15~nm and  630.25~nm lines.
We measure the line ratio of the the pixels inside the orange circle shown in the figure~\ref{fig_mask} (c).
If the intrinsic field strengths are much weaker than 1~kG, 
the Stokes $V$ is proportional to $-g \partial I/ \partial \lambda$,
where $g$ is the Land\'e factor ($g_{6302}=1.667$, $g_{6301}=2.50$) .
In this case, we expect the data points to regress to the dashed line in figure~\ref{fig_ratio}.
The slope of the dashed line $s$ is given by,
\begin{equation}
s=\frac{-g_{6302} \partial I/ \partial \lambda|_{6302}}{-g_{6301} \partial I/ \partial \lambda|_{6301}}=1.69.
\end{equation}
The value of the slope $s=1.69$ is based on the global averages over all the pixels 
in the analyzed subfield of the maximum amplitudes of $\partial I/ \partial \lambda$ in the blue lobe of \ion{Fe}{1} 630.15~nm and  630.25~nm lines.

The data points of the VF pixels inside the HF patches are clearly concentrated along the line 
with the slope of $s=1.69$ (figure~\ref{fig_ratio}~(a)).
Where the VF pixels do not overlap with the HF patches, 
most of data points remain around the dashed line.
However, some points with larger Stokes $V$ amplitudes
start to bend away from this line, and follow the line with the slope of 0.7$s$, shown with the solid line
 (black contours in figure~\ref{fig_ratio}~(b)).
This transition is due to the differential Zeeman saturation for stronger fields between two \ion{Fe}{1} 630.2 lines, 
and  the slope of $0.7s$ indicates the field strength of order of 1~kG \citep{Stenflo2010AA}.

Figure~\ref{fig_ratio}~(c) shows the scatter plot for the network fields indicated with yellow contours
in figure~\ref{fig_mask}~(b). 
The points are completely away from the line with the slope $s=1.69$, 
and regress to the line with the slope of 0.7$s$
(solid line) and 0.6$s$ for higher Stokes $V$ signal.
The coefficients of 0.6 and 0.7 are given by \citet{Stenflo2010AA} for the strong saturated intrinsic field strength.
This scatter plot clearly indicates that the network region consists of strong kG fields.
These strong kG vertical fields located in the internetwork region would be 
formed by convective collapse
\citep{Parker1978ApJ,Hasan1985AA,Nagata2008ApJL,Fischer2009AA,Stenflo2010AA}.

\section{Spatial distribution internetwork magnetic fields on the supergranular scale}
In this section, we qualitatively examine the spatial distributions of 
the internetwork horizontal and vertical magnetic fields on the supergranular scale, 
and aim to address the question of 
whether the  supergranulation affects the distribution of these small-scale internetwork fields.

The area density of the VF patches $d_{\rm VF}(r)$ at a distance $r$ from the center position
of the supergranular cell (cross mark in figure~\ref{fig_mask}~(d))
is given by
\begin{equation}
d_{\rm VF} (r)=\frac{n_{\rm VF}(r)}{n_{\rm all}(r)},
\end{equation}
where, at a distance between $r$ and $r+dr$,
$n_{\rm all}(r)$ is the total number of pixels,
and $n_{\rm VF}(r)$ the number of pixels occupied by the VF patches. 
Here we take $dr=1$~pixel ($\sim0\farcs16$).
When we obtain $n_{\rm all}(r)$ and $n_{\rm VF}(r)$, we do not take into account
the network pixels marked with yellow contours.
Likewise, the area density of HFs  $d_{\rm HF}(r)$ is given by
\begin{equation}
d_{\rm HF}(r)=\frac{n_{\rm HF}(r)}{n_{\rm all}(r)},
\end{equation}
where $n_{\rm HF}(r)$ is the total number of the HF pixels marked 
with red contours at a distance between $r$ and $r+dr$ from the center position of the supergranular cell.

Figure~\ref{fig_plot}~(b) shows the area density $d_{\rm VF}(r)$ and $d_{\rm HF}(r)$.
Five dashed lines indicate $r=4\arcsec, 9\arcsec, 15\arcsec, 22\arcsec,$ and $34\arcsec$, 
and correspond to the orange circle ($r=15\arcsec$) 
in figure~\ref{fig_mask} (c) and four circles in figure~\ref{fig_mask} (d).
We find a significant difference in the spatial distributions  between the HF and VF patches.
The area density of the VF patches, $d_{\rm VF}(r)$, increases with distance.
Meanwhile,  the area density of the HF patches $d_{\rm HF}(r)$ does not change 
from the center to the network boundary,  
and this clearly shows that the HF distribution does not 
have any correlation with the supergranular scale.

\section{Discussion}
\label{conclusion}
\subsection{Spatial coincidence of VF and HF patches}
One of the discoveries  presented in this paper is 
the clear positional association of the VF and HF patches in the internetwork region.
The HF patches are located around the boundary 
of the VF patches as shown in Figure~\ref{fig_plot}~(a), and are connected to the VF patches.
$96\%$ of the total number of the HF patches has the associated VF patches,
and 70\% of the total area occupied by the HF patches overlaps with the VF patches.
Alternatively, 49\% of the total number of the VF patches has the HF patches.
The size of the HF patches are apparently smaller compared 
to that of the VF patches as shown in figure~\ref{fig_mask}~(a).
Only the small core of the HF patches may be seen due to the lower sensitivity 
of the linear polarization to the magnetic field strength.
We quantitatively compare  the individual VF flux with the flux of the corresponding HF patch, 
and find that the VF flux is close to the HF flux (figure~\ref{fig_fluxratio}).
The HF patches so far have been investigated by many authors
as a new magnetic entity 
different from the vertical internetwork fields.
However, they are closely connected,
possibly causally related to the well-known VF patches in the internetwork regions.
The implication of this observational fact will be discussed in the subsequent sections.

As shown in figure~\ref{fig_plot}~(b), 
the area densities of both the HF and VF patches are essentially constant
within $r<15\arcsec$.
This is consistent with the observation that the VF and HF patches are closely connected.
The area density of the HF patches stays the same with radius from the center of the supergranule, 
while that of the VF patches gradually 
increases closer to the supergranular boundary.
This behavior at $r>15\arcsec$ is apparently different from
the correspondence between the HF and VF patches in the pure internetwork region.
Apparent large-size network fields  
indicated with yellow contours in figure~\ref{fig_mask}~(b) are
not included in figure~\ref{fig_plot}~(b).
The histogram of the intrinsic field strengths and the occurrence rates of horizontal fields are 
the same between the quiet Sun and the weak plage region 
when the strong persistent vertical fields (network fields) are removed \citep{Ishikawa2009AA}.    
Furthermore, the histograms of the intrinsic field strengths are 
the same between the quiet Sun near the equatorial limb and the polar region 
after removing the vertical fields including the network fields \citep{Ito2010ApJ}.
The distributions of the HF and VF patches in figure~\ref{fig_plot} (b) are
consistent with the earlier results by \citet{Ishikawa2009AA} and \citet{Ito2010ApJ}.

\subsection{Comparison of longitudinal and transverse apparent flux density}
We estimate and compare the total magnetic flux for both VF and HF patches in the internetwork region.
The mean flux density for unsigned longitudinal (vertical) component 
is estimated to be 8.3~Mx~cm$^{-2}$
and the mean apparent flux density for transverse (horizontal) component is 7.1~Mx~cm$^{-2}$.
We do not find the significant imbalance between them as reported by \citet{Lites2008ApJ}.
As seen in figure~\ref{fig_noise}, we expect that
the photon noise contaminates the real signal within $3\sigma_{\rm tot}$ or $4\sigma_{0}$, 
making it difficult to distinguish the real signals from the noise.
Thus, we exclude these pixels and calculate the mean apparent flux density.
On the other hand, \citet{Lites2008ApJ} used all the pixels for the determination of the
mean apparent flux densities in the FOV; their $1\sigma$ noise levels are 2.4~Mx~cm$^{-2}$ and 41~Mx~cm$^{-2}$ for the longitudinal and transverse apparent flux densities.

The total unsigned vertical magnetic flux in the internetwork region 
defined by the orange circle in figure~\ref{fig_mask}~(c)
is estimated to be 3.0$\times10^{19}$~Mx.
In order to accurately estimate the total magnetic flux density for the HF patches in the internetwork region,
we consider that all the HF patches have the flux tube structure as shown in \citet{Ishikawa2010ApJ}, 
and take into account the cross section of the each horizontal flux tube. 
Indeed, the HF patches are 
considered to be a manifestation of the small-scale bipolar flux tubes 
as discussed at the end of section \ref{property}.

The estimated total magnetic flux of the HF patches 
is 4.7$\times10^{18}$~Mx, which is smaller than that of the VF patches.
One reason for this difference may be different sensitivity 
between Stokes $V$ in detecting the longitudinal (VF) component 
and Stokes $Q/U$ in detecting the transverse (HF) component.
If emerging loops leave the photosphere, there should be more leftover vertical flux, 
which corresponds to the footpoints of these loops, than the horizontal flux.
This is both possible and plausible to have more vertical flux in addition to the sensitivity issue.
Indeed, it is observed that some small-scale loops emerge and reach above the photosphere, leaving behind the bipolar footpoints \citep{Martinez2009ApJ,Ishikawa2010ApJ}.

\subsection{Origin of internetwork fields}
\citet{Isobe2008ApJL} and \citet{Steiner2008ApJL} carried out 
the numerical simulation of the surface layer of the Sun
having an initial condition of vertical unipolar magnetic fields.
They then compared the simulation
with observations.
Their initialization is similar to plage and network regions 
where vertical fields are dominant.
They found that the initially uniform magnetic fields are immediately bent by the convective flows, and
they eventually become disordered due to the convective motion and appear as horizontal fields in the photosphere.
\citet{Steiner2008ApJL} also carried out the simulation initialized with uniform horizontal magnetic fields,
and found that the amount of granular-scale horizontal fields exceeds that of the initialization with uniform unipolar magnetic fields.

We find that the area density of the HF patches does not increase toward the supergranular boundary,
even though the area density of the VF patches increases toward the supergranular boundary.
This independence of the HF patches from the amount of the VF patches indicates that 
the seed fields for the HF patches do not come from the pre-existing vertical magnetic fields, and
our observations appear not to be consistent with the initialization with a unipolar vertical magnetic field that \citet{Isobe2008ApJL} and \citet{Steiner2008ApJL} used.
Although we cannot rule out other possibilities for the origin of the horizontal magnetic fields, it would be conceivable that the granular convective motions amplify the field strength of the seed fields up to the equipartition field strength corresponding to the granular motion 
by stretching and folding,
and drive the emergence of the horizontal magnetic fields from the convection zone.

\subsection{Origin of kG vertical magnetic fields}
\label{originkG}
Measuring the Stokes $V$ line ratio of the \ion{Fe}{1} 630.15/630.25 system \citep{Stenflo2010AA}, 
we obtain the intrinsic field strength in an internetwork region.
The intrinsic field strength inside the HF patches
is dominantly weak ($<1$~kG), and is in the regime of the weak field approximation.
Most of the pixels outside the HF patches 
(but located inside the VF patches)
also have weak intrinsic field strength,
but a significant population of strong (kG) fields may also be seen for those regions (figure~\ref{fig_ratio} (b)).
Such signature of kG intrinsic fields cannot be found inside the HF patches, as clearly shown in figure~\ref{fig_ratio} (a).
The intrinsic field strength in the designated network region is completely above $\sim$1~kG (figure~\ref{fig_ratio}~(c)).

We point out that the observed weak intrinsic field strength inside the HF patches is in the sub-equipartition regime of the magnetic fields strength.
The equipartition field strength corresponding to the granular convection is estimated to be 700~G,
where the density is $\sim10^{-6}$~g~cm$^{-3}$ at a depth of 500~km below the surface and the velocity is 2~km~s$^{-2}$.
On the other hand, some pixels outside the HF patches have field strength above the equipartition value.

As shown in figure~\ref{fig_hist}~(d) the HF patches are statistically located 
inside the granular cells, while the VF patches in the internetwork regions prefer to be located in the darker region, i.e., 
closer to the intergranular lanes, though there is a large spread in intensity.
We conjecture that the magnetic fields below or equal to the equipartition field strength are 
provided by the emergence of granular-scale horizontal fields.
When the emerged flux tubes successfully leave 
the photosphere with left-over footpoints (vertical flux tube), 
the footpoints are intensified beyond the equipartition field strength (super-equipartition field strength of 1-2~kG)
due to convective instability.
We interpret that the scatter plot of figure~\ref{fig_ratio}~(b) shows such evolutionary process from sub-equipartition to super-equipartition fields:
outer higher $\Omega-$loops have the photospheric footpoints (VF patches)
without the HF patches, since the top of such loops 
has already left the photosphere.
We speculate that
such vertical fields corresponding to the VF patches may remain in the sub-equipartition regime, 
or may be subsequently subject to convective collapse, resulting in kG-vertical fields.
Alternatively, the footpoints of the HF patches may almost immediately become strong (super-equipartition)
vertical fields due to convective collapse when the HF patches are advected by the granular flows to the intergranular lanes.

\citet{Nagata2008ApJL} found the observational evidence of the formation of kG flux tube induced 
by convective collapse using SOT/SP \citep[see also ][]{Fischer2009AA};
an equipartition field strength flux tube with the preceding transient downflow of 6~km~s$^{-1}$
goes through the intensification of the field strength to 2 kG. 
The strong downward flow comparable to the local sound speed
is observed in the footpoint region of the small-scale emerging horizontal fields,
suggesting the occurrence of convective collapse 
at the footpoints of horizontal fields \citep{Ishikawa2008AA}.
However, we point out that 
there is no direct piece of evidence that all the kG vertical magnetic fields in the internetwork region 
come from the emergence of horizontal magnetic fields.
Further investigations have to be done.

\subsection{Do horizontal fields concentrate on the supergranular boundary?}
The appearance of the transient horizontal magnetic fields is not spatially random, and has an organized cellular 
structure with a typical scale of 5-10$\arcsec$.
The cellular structures coincide in position
with the negative divergence of the horizontal flow field, i.e.,
mesogranular boundaries with downflows \citep{Ishikawa2010ApJL}. 
As for the internetwork vertical fields, a similar concentration has been observed \citep{Dominguez2003AA}.
Indeed, both the HF and VF patches concentrate in the cellular structure
with a scale of  5-10$\arcsec$ in the snapshot image of figure~\ref{fig_mask}~(a).
These structures would correspond to the mesogranular cells.
A natural question here is whether there is a similar concentration of horizontal magnetic fields in the supergranular boundaries.
We find that the area density of the VF patches increases toward the supergranular boundary,
while the area density of the HF patches is essentially constant with distance. 
The HF patches are not concentrated on the supergranular boundary.

The horizontal flow velocity of the supergranulation at the photosphere is in the range 300-500~m~s$^{-1}$ \citep{Shine2000SoPh}. 
Collapsed vertical fields (Section \ref{originkG}) 
can be advected by the supergraular flow fields, and might form the network fields around the supergranular boundaries.
The abundant vertical fields around the supergranular boundary might be originally provided by the emergence of the horizontal fields in the internetwork region.

\citet{Martin1990} and \citet{Schrijver1997ApJ} 
pointed out that the main source of the network magnetic fields 
are the ephemeral region and that the contribution of the internetwork magnetic fields is not so significant
\citep[see also][]{Wang1987}.
The ephemeral region consists of the emerging bipolar magnetic patches
with a size less than $\sim20\arcsec$.
We find that the internetwork region is filled with 
a lot of small-scale emerging bipolar magnetic fields, 
which are not clearly seen with the previous generation telescope.
The relationship between the ephemeral region and the transient horizontal magnetic fields needs to be clarified.

Here we speculate that vertical magnetic fields in the internetwork fields
are advected by the supergranular horizontal flow and 
that the network fields would be formed.
Using \textit{Hinode}/SOT observation, 
\citet{deWijn2008ApJ} find that apparent flow velocities of magnetic elements (patches) 
in the internetwork region
are slightly biased by 0.2~km~s$^{-1}$ in a radial direction toward the network boundary.
It is clear that the motion of the internetwork magnetic patches is affected by the supergranular flow fields.
However, there is no direct evidence showing that the network magnetic fields are a consequence
of the accumulation of the internetwork magnetic fields.
In order to confirm that the internetwork fields are indeed advected towards 
the supergranular boundaries forming the network fields,
further investigation needs to be done 
with the high spatial and temporal resolution observations using \textit{Hinode}/SOT.

\citet{Ishikawa2010ApJL} point out that the seed fields of the horizontal fields are
advected by the mesogranular flow fields, and 
granular convective motion amplifies the field strength from below the mesogranular 
equipartition field strength $B_{\rm eq}$ to granular $B_{\rm eq}$,
and drives the emergence of the horizontal fields. 
If the seed fields are advected by the mesogranular flow fields,
the seed fields are also likely to be advected by the supergranular flow:
we should have seen the concentration of the horizontal fields around the supergranular boundaries.
We do not have a clear interpretation for this asymmetry between mesogranule and supergranule
in terms of the population of the HF patches. 
This issue would be related to a fundamental relationship 
between the small-scale magnetic fields and the convection with different scale sizes.

\section{Summary and Speculation}
In the present paper,
we have analyzed the high spatial resolution spectropolarimetric data obtained with the SOT/SP
to investigate the properties of the internetwork magnetic fields. 
We have paid special attention to the effect of the photon noise in the analyzed data, 
and chosen the valid pixels minimally contaminated by the noise.
Therefore, the present paper is limited in scope to the field with sufficiently strong polarization signal.
An obvious penalty of having the stringent noise criteria is that we do not detect weak magnetic fields which provide polarization signal
comparable to or weaker than the noise level.
We do not rule out the existence of such weak fields.
We refer the pixels with the significant Stokes $V$ signal to the VF (vertical field) pixels, and
those with the significant Stokes $Q$ or $U$ to the HF (horizontal field) pixels.
The main observational results with our immediate interpretation are summarized as follows:
\begin{enumerate}
\item We find positional association 
between the vertical and the horizontal magnetic fields in the internetwork region.
The vertical magnetic fields correspond to the weak vertical magnetic fields 
known for decades to exist.
The horizontal magnetic fields are investigated in detail with \textit{Hinode} for the first time.
Approximately half of the total number of the VF patches 
accommodates the HF patches partially or entirely, and 
essentially all of the HF patches overlap with the VF patches.
These HF patches are typically located around the edge of the VF patches.
We compare the VF flux with the flux of the corresponding HF patch,
and find that HF flux is close to the VF flux.
These observations suggest that the HF patches are physically and causally connected to the VF patches in the internetwork region.
It has been discussed that the transient horizontal magnetic fields are a manifestation
of the local dynamo process \citep{Ishikawa2009AA,Ito2010ApJ}.
The origin of the vertical magnetic fields in the internetwork region is not yet known,
and this positional coincidence may allow us to 
reach a unified understanding of the internetwork fields in general.

\item Approximately half of the HF patches have bipolar patches (Stokes $V$ signals),
and the other half have the unipolar Stokes $V$ signal.
The unipolar HF patches have both a smaller size and smaller horizontal magnetic flux densities.
We suggest that only dominant polarity is seen
due to the cancellation of the Stokes $V$ signal within the resolution element 
in the unipolar HF patches,
and that the HF patches have an essentially bipolar structure.
Another possibility is that a small portion of the bipolar HF patches may be detected as unipolar HF patches.

\item The mean longitudinal (vertical) magnetic flux density is 
estimated to be 8.3~Mx~cm$^{-2}$, and the apparent transverse (horizontal) one is 7.1~Mx~cm$^{-2}$.
The former number is close to the one derived in the previous studies 
\citep[e.g.,][]{Orozco2007ApJL,Lites2008ApJ}, 
but the apparent transverse flux density derived here is a factor of 5 smaller than the one obtained by \citet{Lites2008ApJ}. 
In our analysis, the mean magnetic flux density of the vertical fields is almost the same as that of the horizontal fields.
We also estimate the total magnetic flux of the HF patches, taking into account the cross section
of the horizontal flux tubes,
and find that the total magnetic flux of the HF patches is a factor of 3 smaller than that of the VF patches.
This excess of the total magnetic flux of the VF patches would indicate that the some portion of the vertical fields 
inside the VF patches reach above the photosphere.

\item Measuring the Stokes $V$ polarization of the \ion{Fe}{1} 630.15-630.25~nm line system
(line ratio method; \citet{Stenflo1973SoPh}), 
we infer the intrinsic field strength in the internetwork region.
The intrinsic field strengths inside the HF patches (overlapping area of the VF and HF patches) 
are weak ($<1$~kG), 
while the strengths outside the HF patches (non-overlapping area) range from weak to strong (kG). 
This would suggest that convective collapse takes place at the footpoint (vertical portion) 
of the emerging horizontal field that is about to leave or that has left the line-forming region in the photosphere,
and this is considered to be one of the mechanisms to produce strong (kG) fields in the internetwork region.
The two distinct flux populations represent the pre-collapsed and the post-collapsed vertical flux tubes
\citep{Stenflo2010AA}.
The snapshot image analyzed in this paper contains the two evolutionary states.

\item It is well known that the VF patches concentrate 
on the mesogranular and supergranular boundaries, 
while we show in this paper that the HF patches concentrate only on the mesogranular boundaries of the supergranunlar cell studied here.
Only the VF patches concentrate around the supergranular boundary.
In the internetwork region within $15\arcsec$ from the center position of the supergranulation,
the area densities of both VF and HF patches are constant with radius.
The area density of only VF patches increases
with increasing radius outside $15\arcsec$ radius (closer to the network boundary).
The area density of HF patches does not increase even though
there are more VF patches closer to the supergranular boundary,
and this independence of the amount of the HF patches
suggests that the HF patches are not due to the bending of pre-existing vertical magnetic fields
with the granular convective motion \citep{Isobe2008ApJL}.
 \end{enumerate}
 
These observations shed light on the origins of both the vertical and horizontal magnetic fields
and their causal relationship, potentially leading to a new picture on the quiet Sun
magnetism.
As a first step for a unified understanding on the quiet Sun magnetism,
we here proposed one possible scenario for generation and maintenance of  
the internetwork vertical and horizontal magnetic fields:
(1) mesogranular flow fields may advect the seed fields with field strength smaller than
mesogranular $B_{\rm eq}$ into the mesogranular boundaries.
The granular convective motion amplifies the field strength of 
the seed fields up to granular $B_{\rm eq}$ by folding and stretching, 
and the granular upward motion carries them to the photosphere.
These are observed as the small-scale horizontal magnetic fields on the photosphere.
This process is discussed in \citet{Ishikawa2010ApJL}.
(2) Some horizontal fields reach above the photosphere, leaving their footpoints (bipolar magnetic fields).
(3) Vertical magnetic fields, which are the footpoints of the emerged 
small-scale loops, are advected 
toward the intergranular lane, and are subject to convective collapse, partially forming strong kG fields.
(4) Finally, these vertical fields are transported to the supergranular boundaries, and form the network magnetic fields.

The occurrence rate of the emergence of the horizontal magnetic fields is 
quite high \citep{Ishikawa2009AA}.
It is likely that such horizontal fields do not return to the photosphere 
and that they reach the chromosphere and possibly corona \citep{Martinez2009ApJ,Ishikawa2010ApJ}.
It is found that 23\% of the small-scale loops on the photosphere
reach above the line forming region of \ion{Fe}{1}~630.2~nm lines \citep{Martinez2009ApJ},
and the total magnetic flux carried by the horizontal magnetic fields above the photosphere
is estimated to be two orders of magnitude larger than the total magnetic flux of the sunspots \citep{Jin2009ApJ,Ishikawa2010ApJ}.
The rest of them would be shredded, 
thus falling below the threshold of detectability,
and/or return below the photosphere.
It is plausible that the emergence of these ubiquitous small-scale loops leaves the vertical magnetic fields on the photosphere, and is the source of the quiet Sun magnetic fields with the process suggested above.

\acknowledgements 
The idea of this work was inspired by the discussions in 
the 6th Solar Polarization Workshop on 2010 held in Maui.
\textit{Hinode} is a Japanese mission developed and launched by ISAS/JAXA, with NAOJ as a domestic partner and NASA and STFC (UK) as international partners. It is operated by these agencies in co-operation with ESA and NSC (Norway).

\end{document}